\documentclass[prd,aps,twocolumn,amsmath,amssymb,floatfix,superscriptaddress]{revtex4}
\usepackage[dvips]{graphicx}
\begin{document}

\title{Variability of signal to noise ratio and the network analysis of gravitational wave burst signals }

\author{Soumya D. Mohanty}
\email{mohanty@phys.utb.edu}
\affiliation{The University of Texas at Brownsville, 80 Fort Brown, 
Brownsville, Texas, 78520, USA}
\author{Malik Rakhmanov}
\altaffiliation[Currently at: ]{Department of Physics, Pennsylvania State University, University Park, PA 16802, USA}
\author{Sergei Klimenko}
\author{Guenakh Mitselmakher}
\affiliation{University of Florida, P.O.Box 118440, Gainesville, Florida, 32611, USA}

\begin{abstract}
%focus more on this work than constraint likelihood 
The detection and estimation of gravitational wave burst signals, with {\em a priori} unknown polarization waveforms, 
requires the use of data from a network of detectors. For determining how the data from such a network should be combined,
approaches based on the maximum likelihood principle have proven to be useful.
The most straightforward among these uses the global maximum of the likelihood over the
space of all waveforms as both the detection statistic and signal estimator.
However, in the case of burst signals, a physically counterintuitive situation results: for two aligned detectors
the statistic  includes the cross-correlation of the detector outputs, as expected,
but this term disappears even  for an infinitesimal misalignment. This {\em two detector paradox}
arises from the inclusion of improbable waveforms in the solution space of maximization. Such waveforms 
produce widely different responses in detectors that are closely aligned.
We show that by penalizing waveforms that exhibit large
  signal-to-noise ratio (snr) variability, as the corresponding source is moved on the sky, 
 a physically motivated restriction is obtained  that (i) resolves the two detector paradox and (ii) leads to
 a better performing statistic than the global maximum of the likelihood. 
Waveforms with high snr variability turn out to be precisely the ones that are
 improbable in the sense mentioned above.  The coherent network analysis method thus obtained can be 
applied to any network, irrespective of the number or the mutual alignment of detectors.
\end{abstract}

\maketitle

\section{Introduction}
\label{intro}
Several interferometric gravitational wave (GW)  detectors are now operational around the world~\cite{LIGO,VIRGO,GEO600,TAMA} in addition to an already operating set of resonant mass detectors~\cite{IGEC}. Since gravitational wave detectors can measure both the amplitude and phase of incoming signals, it is possible to combine the data from such a network of detectors to not only detect GW sources better, compared to single detectors, but also estimate the waveforms of the two independent polarization components of the GW signal along with the direction to the source.  

%join with above
A straightforward approach to signal detection and estimation using data from a network of detectors is 
%by means of
through the likelihood~\cite{Stuart+Ord:v2} functional.  
The likelihood functional of data ${\bf x}$ is the joint probability density function $p({\bf x}|{\bf \theta})$, where
 $\theta$ is a vector of parameters whose true value for any given instance of ${\bf x}$ is not known {\em a priori}. In the case of GW detection, ${\bf x}$ is the collection of outputs from all the detectors in a network and the parameters are the sky position of the GW source and the waveforms
of the two GW polarization components, $h_+(t)$ and $h_\times (t)$. The problem of signal detection is concerned with deciding whether ${\bf x}$ is a realization of $p({\bf x}|{\bf \theta}_0)$ or $p({\bf x}|{\bf \theta}\neq{\bf \theta}_0)$, 
where ${\bf \theta}_0$ denotes the signal absent case. 
The estimation problem deals with infering the true value of ${\bf \theta}$ once a detection has been made. According to the maximum likelihood~\cite{Stuart+Ord:v2} approach to inference, the value ${\bf \theta}_{\rm max}$ at which the likelihood functional attains its maximum
 serves as an estimate of the true value of ${\bf \theta}$. The maximum, over ${\bf \theta}$, of the likelihood  ratio (LR)
 functional $p({\bf x}|{\bf \theta})/p({\bf x}|{\bf \theta}_0)$ yields a detection statistic. Thus, the likelihood based approach conveniently yields both a detection statistic and source parameter estimates.

For binary inspiral signals, the polarization waveforms can be computed in terms of a small set of parameters describing the binary, such as the masses of the two stars, the orbital inclination etc. Likelihood based network analysis is fairly well understood in the case of such known waveforms~\cite{finn,bose+etal}. 
%A similar
% situation prevails for a stochastic GW background~\cite{stochastic} signal. 
In this paper, we consider likelihood based
network analysis for burst signals that are short duration deterministic signals but have
 unreliable or simply unknown waveforms.  Astrophysical sources of burst signals include Supernovae~\cite{corecollapse}, Gamma Ray Bursts~\cite{Kobayashi+Meszaros} and the merger of
compact objects such as Neutron stars and Black Holes~\cite{flanagan+hughes:I}.

Although we do not have prior knowledge of burst signal waveforms, it is possible to extend
likelihood based network analysis to burst signals in a formal sense. The basic 
%sort references
idea~\cite {Flanagan+Hughes:II,mohanty+etal:gwdaw8,johnston,Klimenko+etal:2005} is to treat each sample of a burst signal as a parameter 
and maximize the likelihood over all the parameters. 
We refer to the resulting detection method as  standard likelihood (SL).
Other approaches to network analysis for burst signals have also been proposed in the literature~\cite{guersel+tinto,sylvestre}. The method~\cite{guersel+tinto} proposed by G\"ursel and Tinto is based on extremizing a functional that is constructed purely out of the response of each detector in a network and does not explicitly depend on the polarization waveforms.  A minimum of three misaligned 
detectors are required in order to use the G\"ursel-Tinto method whereas likelihood based methods do not have this limitation. 

Although it is quite elegant, the SL method has a fundamental problem that was first noted in~\cite{johnston} and systematically analyzed in~\cite{Klimenko+etal:2005} (henceforth, KMRM). 
Called the two detector paradox, the problem is most easily stated for a network
of two detectors. When the detectors are perfectly aligned,  the SL detection statistic naturally 
includes the cross-correlation of detector outputs. Since the response to a given GW source is identical 
in aligned detectors, this term serves to increase the sensitivity of the overall detection statistic. However, it disappears when the SL approach is used for even infinitesimally misaligned detectors. This is counterintuitive since we expect the cross-correlation term to remain important for small misalignments.

  The origin of the two detector paradox is that all waveforms for $h_+(t)$ and $h_\times (t)$ are allowed as solutions
when maximizing the likelihood even though some of them would be improbable
 in that they will produce widely different responses in two nearly aligned detectors. 
It follows that this problem can be solved by removing such improbable waveforms from the space of allowed solutions.
In KMRM, this was achieved by imposing certain constraints on the space of waveforms yielding the class of constraint likelihood methods. 
 Modifying the likelihood functional with a waveform dependent weight factor, such as a Bayesian prior~\cite{bayes}, is another
way of suppressing undesirable solutions. 
In the following we use the term ``restriction"  to encompass all approaches that remove improbable solutions or reduce their influence.

%make it easier to read
The non-trivial aspect of setting up restrictions is that the performance of
the resulting statistic must not be degraded and, if possible, should be better than the SL approach. 
Additionally, the restrictions must be applicable to all astrophysical burst sources otherwise the resulting 
statistic will be better for a particular type of source but may be worse for others.  
%Indeed, the existence of the two detector paradox itself points to the possibility of formulating restrictions with the said property. 
%Note that the converse of the above, namely, that a constraint free of source dynamics should lead to better performance, need not be true. This must be %verified on a case-by-case basis.
If sufficiently reliable knowledge for a particular class is available, then it can be used~\cite{Anderson+etal,Brady+Mazumdar} as 
an additional set of restrictions to further enhance sensitivity. 
%An extreme example is binary inspirals where our knowledge of waveforms is reliable enough to reduce the likelihood maximization problem to matched %filtering. 
The constraints proposed in KMRM (see Section~\ref{2dp}) satisfy the above condition, namely independence from assumptions about source 
dynamics. Simulations show that constraint likelihood methods perform significantly better than SL, especially in reconstructing the sky position of a burst source. However, the constraints used were a first example and are not necessarily unique or optimal. 
It is important to investigate independent avenues of formulating generic restrictions 
in order to gain more insight into this promising approach to network analysis. 

We investigate the issue of generic restrictions from a physical point of view. 
%Superficially, it resembles the Bayesian approach but our ``prior" is not normalized and the detection statistic is a secondary maximum of the Likelihood %Ratio instead of the posterior probability distribution.
It is found that the more improbable a set of waveforms $h_+$ and $h_\times$, the faster is the change in detectability of the corresponding source
 as a function of sky position. By measuring the rate of change of signal-to-noise ratio (SNR) as a function of sky position, therefore, we can restrict
 solutions that are undesireable in the maximization of the LR. 
%We find that sources that produce widely different responses in nearly aligned detectors also have high
% SNR variability and, thus, are unlikely to be %detected often. 
%The measure is  used as a 
%penalty term in the maximization of the LR over the space of waveforms. 
%Our approach sacrifices detectability of such sources in order to boost
% sensitivity  to ``normal" sources.
 Monte Carlo simulations show that the detection statistic that results from the above 
restriction performs roughly the same as constraint likelihood methods. However, there exist important qualitative differences in how the
two methods process data. For example, unlike constraint likelihood, the divergence from SL of the
method introduced here depends on the actual data.

The paper is organized as follows. We review the standard likelihood method in Section~\ref{netanalysis}, establishing our basic notation and conventions on the way. Section~\ref{2dp} contains a discussion of the two detector paradox. Section~\ref{snrvar} considers the SNR variability of signals, discusses how it affects their detectability and defines a measure for it. Section~\ref{penlike} contains the derivation of the detection statistic and source parameter estimators that result from incorporating SNR variability. Results from simulations are presented in Section~\ref{results}.
%%%%%%%%%%%%%%%%%%%%%%%%%%%%%%%%%5
\section{Standard likelihood analysis for burst signals}
\label{netanalysis}
We fix a Cartesian coordinate system with its origin at the center of the Earth, $Z$ axis pointing at the North Pole and some arbitrary convention for the $X$ and $Y$ axes. This will be our fiducial frame of reference. Let the sky position of an incoming GW signal in this frame be denoted by $\theta$, the polar angle, and $\phi$, the azimuthal angle. An incoming GW signal is most conveniently expressed in the transverse traceless (TT) guage~\cite{MTW} associated with the source direction.  The $Z$ axis of the TT frame is oriented along the direction to the GW source. We adopt the convention that the $X$ axis of the TT gauge lies in the plane formed by the TT frame $Z$ axis and the $Z$ axis of the fiducial frame. The GW signal is specified by its two independent polarization components $h_+(t)$ and $h_\times (t)$ in the TT frame.  

%Soften this... we do not take into account...
In this paper, we take into account the  different orientations and geographic locations of currently operating interferometric detectors
but treat them as identical in sensitivity.  
%In particular, when obtaining numerical results, 
%the detectors will be assumed to have equal armlengths (say, 4~km) but will have the same locations and orientations as the real detectors
%currently in operation.
 For instance, a GEO detector in the paper will mean a  detector having the same location and orientation as GEO600
but with sensitivity similar to that of LIGO detectors. 
This simplifies the algebra considerably whithout altering the main result
which is the relative comparison of methods.

%get rid of W
%reference kip thorne 
Since real data is digital, we shift to the discrete time domain and from here on denote any time dependent quantity $g(t)$ as $g[j]$, where $t=j\delta$, $\delta$ being the sampling interval. 
A finite sequence of time samples $\{g[0],g[1],\ldots,g[m-1]\}$ will be denoted by $\overline{g}$.
The scalar response of a detector $s[k]$ to an incoming 
GW is given by,
\begin{equation}
s[k] = W^{ij}[k] D_{ij}\;,
\end{equation}
where $W^{ij}$ is the symmetric trace free tensor that describes the GW in the transverse-traceless gauge and $D_{ij}$ is an equivalent tensor associated with the detector~\cite{dhurandhar+tinto}. (The Einstein summation convention is in force in the above expression.) 
%Both tensors can be transformed into the common fiducial frame via rigid rotations. 
The response $s_i[k]$ of the $i^{\rm th}$ detector, $i=1,\ldots,N_d$, is expressed as,
\begin{equation}
s_i[k] = F_{+,i}(\theta,\phi)h_+[k\delta-\delta_i] + F_{\times,i}(\theta,\phi) h_\times[k\delta-\delta_i]\;,
\end{equation}
where $F_{+,i}$ and $F_{\times,i}$ depend on the rotations involved in going from the TT frame associated with a detector to the wave frame
 associated with the fiducial frame and $\delta_i$ is the travel time for the signal from the origin of the fiducial frame to the detector. 
For a given $\theta$, $\phi$, we can treat the detectors as co-located ($\delta_i=0$) at the origin of the fiducial frame since the output from each detector can be shifted in time to compensate for $\delta_i$. 

For a network of detectors, 
it is convenient to define quantities that are analogous to those for a single detector. Thus,
 we can define a network response vector,
\begin{equation}
{\bf S}[k] = \left(
\begin{array}{c}
s_1[k]\\
\vdots\\
s_{N_d}[k]
\end{array}
\right)\;.
\end{equation}
Vectors in $\mathbb{R}^{\rm N_d}$, the space of real $N_d$-tuples, will be denoted by boldface capital letters.
 By convention,  every vector ${\bf A}\in \mathbb{R}^{\rm N_d}$ will be a column vector. We define
the Euclidean scalar product  in this space, ${\bf A}{\cdot}{\bf B}={\bf A}^T{\bf B}$. 
A sequence of vectors will be denoted by $\overline{\bf A} = \{ {\bf A}[0],\ldots,{\bf A}[m-1]\}$.

For given $\theta$, $\phi$,
\begin{equation}
\label{ndetresponse}
{\bf S}[k] = {\cal F}(\theta,\phi) \left(
\begin{array}{c}
h_+[k]\\
h_\times[k]
\end{array}
\right)\;,
\end{equation}
where the matrix ${\cal F}$ is defined as,
\begin{equation}
{\cal F}(\theta,\phi) = \left(
\begin{array}{cc}
F_{+,1}(\theta,\phi) & F_{\times,1}(\theta,\phi)\\
\vdots & \vdots\\
F_{+,N_d}(\theta,\phi) & F_{\times,N_d}(\theta,\phi)
\end{array}
\right)\;.
\end{equation}
Given a network response vector ${\bf S}[k]$, the polarization components can be obtained as,
\begin{eqnarray}
\left(\begin{array}{c}h_+[k]\\ h_\times[k] \end{array}\right) &=&  {\cal H}^{-1}{\cal F}^T(\theta,\phi) {\bf S}[k]\;,\\
{\cal H}(\theta,\phi) &=& {\cal F}^T(\theta,\phi){\cal F}(\theta,\phi)\;,\label{def_calh}
\end{eqnarray}
provided ${\cal H}^{-1}$ exists.
 (Note that ${\cal H}$ is a symmetric 2-by-2 matrix and, hence, has two real eignevalues.)

Let the noise in the $i^{\rm th}$ detector be $\overline{n}_i$. In the following we will assume 
that $\overline{n}_i$ is a zero mean, white, Gaussian process, i.e.,  ${\rm E}[n_i[j]]=0$ and 
${\rm E}[n_i[j] n_m[k]]=\delta_{im}\delta_{jk}$. In analogy to the network response vector ${\bf S}[k]$,
 we define a network data vector ${\bf X}[k]$,
\begin{equation}
{\bf X}[k] = \left(
\begin{array}{c}
x_1[k]\\
\vdots\\
x_{N_d}[k]
\end{array}
\right) = \left(\begin{array}{c} n_1[k]\\ \vdots \\ n_{N_d}[k]\end{array}\right) + {\bf S}[k]\;.
\end{equation}
The natural logarithm of the likelihood ratio (LR), $\Lambda(\overline{\bf X}|\theta,\phi,\overline{\bf S})$, for an $N_d$ detector network  is given by
 \begin{eqnarray}
\Lambda(\overline{\bf X}|\theta,\phi,\overline{\bf S}) &=&\frac{1}{2} \sum_{i=1}^{N_d}\! \sum_{j=0}^{N-1}\! \left(x_i[j]^2-(x_i[j]-s_i[j])^2\right) \\
&=& \sum_{j=0}^{N-1}\left[ {\bf  X}[j]{\cdot}{\bf  S}[j] - \frac{1}{2}\left\|{\bf  S}[j]\right\|^2\right]\;,
\label{loglratio}
\end{eqnarray} 
where the detectors outputs are implicitly time shifted as explained above.
 
The maximum likelihood ratio approach can be extended to burst signals by substituting Eq.~\ref{ndetresponse} into Eq.~\ref{loglratio} and maximizing $\Lambda(\overline{\bf X}|\theta,\phi,\overline{\bf S})$ over each $h_+[j]$, $h_\times[j]$ independently. Since the resulting equations involve only quantities at the same sample $j$, this involves maximizing the summand in Eq.~\ref{loglratio} independently for each $j$. 
%As a result, a simple geometrical picture emerges. 
From now onwards we will focus on only one time instant and drop the time index when there is no scope for confusion.
The quantity to be maximized is,
\begin{equation}
\lambda[j] = {\bf  X}[j]{\cdot}{\bf  S}[j] - \frac{1}{2}\|{\bf  S}[j]\|^2\;,
\label{llrsummand}
\end{equation}
over the vector ${\bf  S}$ in $\mathbb{R}^{N_d}$, the space of real $N_d$-tuples. Let the maximum of $\lambda$ occur at ${\bf  S}_{max}$. If the maximization allows all possible ${\bf  S}$, then it is obvious that ${\bf  S}_{max} ={\bf  X}$. However, we are not allowed to choose any arbitrary ${\bf  S}$ in the evaluation of $\lambda$ since from Eq.~\ref{ndetresponse}, ${\bf  S}$ is a vector that depends on only two parameters $h_+[j]$ and $h_\times[j]$. This implies that ${\bf  S}$ must lie in a two dimensional plane determined completely by the antenna pattern functions of the detectors and passing through the origin of $\mathbb{R}^{N_d}$. 
%Figure~\ref{fig1} reinforces this statement with a trivial illustration. In the following, we call the two dimensional 
%plane to which all detector responses are confined as the {\em response plane}. 

Though the maximization of $\lambda$ is restricted to $\bf S$ lying in the response plane, in the maximum likelihood ratio approach there are no further constraints on the choice of $\bf S$ within this plane. We call the method of maximizing $\lambda$ over ${\bf S}$ with no further restrictions as the standard likelihood (SL) method~\cite{flanagan+hughes:I,mohanty+etal:gwdaw8,johnston,Anderson+etal,Klimenko+etal:2005}. We denote the value of  $\Lambda$ (Eq.~\ref{loglratio}), obtained by summing the maximized values of $\lambda$ corresponding to each time instant, by $\Lambda_{SL}(\theta,\phi)$. This is the detection statistic of the SL method. 
%\begin{figure}
%\includegraphics[height=2.5in,width=3.0in]{snrgrad_fig1.eps}
%\caption{Maximization of $\lambda$ defined in Eq.~\ref{llrsummand}. The data vector ${\bf  X}$ lies in $R^{N_d}$ while the response vector over which
%$\lambda$ is maximized
% is constrained to a two dimensional plane. The orientation of the plane is fixed by the detector antenna patterns.
%\label{fig1}}
%\end{figure}
%%%%%%%%%%%%%%%%%%%%%%%%%%%%%%%%%
\section{The two detector paradox}
\label{2dp}

In the two detector case, the network data vector ${\bf X}$ lies in the response plane. 
When the detectors are aligned, we are constrained to use a network response vector ${\bf  S}$ lying along the $45^\circ$ line since
the responses of the two detectors must be identical, $s_1[k]=s_2[k]=s[k]$ for all $k$. Maximizing $\lambda$ over the response $s[k]$
yields,
\begin{eqnarray}
\widehat{s} &=& \frac{1}{2}(x_1+x_2)\;,\\
\widehat{\lambda} &\propto& \frac{1}{2}(x_1+x_2)^2 \;,
\end{eqnarray}
where $\widehat{\lambda}$ denotes the maximum value and $\widehat{\lambda} =\lambda(\widehat{s})$.
 It follows that,
\begin{equation}
\Lambda_{SL}(\theta,\phi) \propto \sum_{j=0}^{N-1} \left(x_1[j]^2 + x_2[j]^2 + 2 x_1[j] x_2[j]\right)\;.
\end{equation}
 Thus, the SL detection statistic contains a cross-correlation term, $\sum x_1[j] x_2[j]$, which agrees with
our intuition: since the responses of the two detectors are identical, the cross-correlation term will always have a positive definite mean
in the presence of a signal and, therefore, should be useful for detection. 

However, the same procedure, when carried out for even infinitesimally misaligned detectors, yields $\Lambda_{SL}$ that has
no cross-correlation term,
\begin{eqnarray}
\widehat{s}_1 = x_1 & ;& \widehat{s}_2 = x_2\;,\\
\widehat{\lambda} &\propto& x_1^2 + x_2^2\;.
\end{eqnarray}
 This clearly violates what one expects on physical grounds  since the cross-correlation term
should remain a useful discriminant against noise when the detectors are only infinitesimally misaligned. 
However, for the class of probable waveforms, the SL statistic becomes less sensitive than one in 
which a cross-correlation term is added by hand.
Thus, the SL
 approach fails to deliver a physically acceptable detection statistic in the two detector case. Following KMRM we call  this problem
 the two detector paradox. 

The origin of the two detector paradox lies in the maximization of $\lambda$ over  all waveforms, $h_+$ and  $h_\times$, 
including those
that can produce very different responses $s_1$ and $s_2$ for even infinitesimally misaligned detectors. Such
waveforms can lead to small or even zero cross-correlation between the responses. Physical intuition suggests that such waveforms should 
 be imrobable. Mathematically, however, such waveforms are allowed since the matrix ${\cal F}$ is invertible, even if it is nearly singular, and some $h_+$ and $h_\times$ can always be obtained for any given $s_1$ and $s_2$. As a result, the SL
 approach ``throws" out the cross-correlation term since this term can now contribute pure noise.
 Geometrically, the two detector paradox comes about because the data vector ${\bf X}$ lies in the response plane and the SL
 approach does not constrain the choice of ${\bf S}$ in this plane. Hence, the solution $\widehat{\bf  S}={\bf X}$ that removes cross-correlation is an allowed one in this case. This suggests that restricting the allowed solutions in the response plane can resolve the two detector paradox. The SL
 detection statistic recovers pair-wise cross-correlations in the case of three or more misaligned detectors since ${\bf X}$ now lies outside the response plane. However, the two detector paradox is just an extreme manifestation of a 
problem that occurs for any arbitrary network: depending on the geometry of the network, some solutions are improbable and their inclusion in the 
maximization of the LR reduces its sensitivity. 

\section{Variability of SNR over the sky} 
\label{snrvar}
Though our physical intuition suggests that an infinitesimal misalignment between two detectors should not 
result in very different responses $s_1 \simeq s_2$, this does not forbid the existence of such sources. As discussed above,
mathematically, $h_+(t)$ and $h_\times (t)$ can be completely arbitrary and, in particular, can be such that $s_1$ is very different from $s_2$. 
However,  we expect  such sources to be improbable. Is their a quantitative basis for our expectation? In this section we show that, indeed, there is a natural argument that ``probable" GW sources, for which $s_1\simeq s_2$,  should occur more often than ``improbable"
 ones. Note that the categorization of
a source as probable or improbable also depends on the mutual alignment of detectors. For example, 
two detectors oriented at $45^\circ$ to each other can often have
very different responses for the same source. As shown below, our argument takes this factor into account.  
We first present the two detector case and then consider an arbitrary network.
%We propose what appears to be a natural measure for distinguishing between different choices of ${\bf  S}$. As we will see, this measure
% forces the orientation of ${\bf  S}$ to be
% close to the $45^\circ$ line when the detectors are closely aligned but allows ${\bf  S}$ to be arbitrarily oriented
%when the misalignment is large. Hence, maximizing $\lambda$ over ${\bf S}$ while keeping
%track of this measure resolves the two detector paradox. 

\subsection{Two detector case}
%%simplify
\label{snrvar2det}
%Let there be two observers A and B who are observing GWs with their own pairs of GW detectors $D_1^{a}$, $D_2^{a}$ and $ D_1^{b}$, $D_2^{b}$ %respectively. Let  the  detectors for observer B be such that they can be obtained by taking the fiducial frame, containing only the detectors of observer A, 
%and rotating it rigidly. 
%See Fig~\ref{fig2}
%for an illustration.
%\begin{figure}
%\includegraphics[height=2.0in,width=2.5in]{snrgrad_fig2}
%\caption{\label{fig2}
%B and A's detector pairs. A's pair is shown as solid arrows (the interferometers are assumed to be have perpendicular arms). B's pair is the
%dashed set of arrows. For simplicity, $D_1^{a}$, $D_2^{a}$ are shown as coplanar but misaligned with respect to each other.
%Similarly $ D_1^{b}$ and $D_2^{b}$ are coplanar and misaligned with respect to each other by the same amount. The corresponding planes in which the %detectors lie are shown in the respective styles (solid: A, dashed: B) of the
%detector arrows.
%}
%\end{figure}
Consider a gravitational wave source located at $(\theta, \phi)$ with signal polarization components $h_+[k]$ and $h_\times[k]$. Let the detector response vector  be ${\bf S}[k]$. Now consider an identical source located at $(\theta^\prime,\phi^\prime)$. By this we mean that the fiducial frame, along with the GW source,  be rotated rigidly leaving the detectors in their original place. This leads to a  network response ${\bf S}^\prime[k]$ 
that is related to ${\bf S}[k]$ by,
\begin{equation}
{\bf S}^\prime = {\cal F}(\theta^\prime,\phi^\prime){\cal F}^{-1}(\theta,\phi) {\bf S}\;.
\end{equation}
Now, assume that we use matched filtering~\cite{helstrom} to detect the source at the instant $k$. Thus, our detection statistic will be ${\bf X}[k].{\bf S}[k]$
and ${\bf X}[k].{\bf S}^\prime[k]$ for the two sources respectively. The detectability of a source under matched filtering is measured by the  signal to noise ratio (SNR), $\rho$, defined as the ratio of the mean of the detection statistic to its standard deviation. For the two sources above, the respective SNR will be $\rho=\|{\bf S}[k]\|$ and $\rho^\prime=\|{\bf S}^\prime[k]\|$. Note that we are considering a more restricted version of matched filtering than the conventional one in which the detection statistic would be $\sum {\bf X}[k].{\bf S}[k]$ and the SNR would be $\rho^2= \sum \|{\bf S}[k]\|^2$. We return to this point later in this section. 

 We denote the relative difference in SNR for the two source locations as,
\begin{eqnarray}
\epsilon(\theta,\phi,\theta^\prime,\phi^\prime;\widehat{S}) &=& \frac{\rho^\prime-\rho}{\rho}\;,
%\\
%&=& ||{\cal F}(\widetilde{\theta},\widetilde{\phi}){\cal F}^{-1}(\theta,\phi) {\bf S}||/||S||-1\;,\nonumber\\
%&=& ||{\cal F}(\widetilde{\theta},\widetilde{\phi}){\cal F}^{-1}(\theta,\phi) \widehat{S}||-1\;,
\label{snrvardef}
\end{eqnarray}
where $\widehat{S}$ is the unit vector oriented along ${\bf S}$. For small displacements, 
$\theta^\prime=
\theta+\delta\theta$ and $\phi^\prime=\phi+\delta\phi$,
 we get,
%\begin{widetext}
\begin{eqnarray}
\epsilon(\theta,\phi,\theta^\prime,\phi^\prime; \widehat{S}) & = &
\delta \theta \left.  \frac{\partial \epsilon}{\partial\theta^\prime}\right|_{\substack{\theta^\prime=\theta,\\ \phi^\prime=\phi}}
+\delta\phi \left. \frac{\partial\epsilon}{\partial \phi^\prime}\right|_{\substack{\theta^\prime=\theta,\\ \phi^\prime=\phi}}\;,\\
\left.  \frac{\partial \epsilon}{\partial\theta^\prime}\right|_{\substack{\theta^\prime=\theta,\\ \phi^\prime=\phi}}
&=& \left.  \frac{\partial}{\partial\theta^\prime} \sqrt{\| {\cal F}(\theta^\prime,\phi^\prime) {\cal F}^{-1}(\theta,\phi) \widehat{S}\|^2}
\right|_{\substack{\theta^\prime=\theta,\\ \phi^\prime=\phi}}
\;,\nonumber\\
%&=& \left.  \frac{1}
%{2|| {\cal F}(\widetilde{\theta},\widetilde{\phi}) {\cal F}^{-1}(\theta,\phi) \widehat{S}||}
% \frac{\partial}{\partial \widetilde{\theta}}\left( \widehat{S}^T {\cal F}^{-1}(\theta,\phi)^T {\cal F}^T(\widetilde{\theta},\widetilde{\phi})
%{\cal F}(\widetilde{\theta},\widetilde{\phi}) {\cal F}^{-1}(\theta,\phi) \widehat{S}\right)
%\right|_{\widetilde{\theta}=\theta, \widetilde{\phi}=\phi}\;,\nonumber\\
&=& \widehat{S}\cdot \left(\frac{\partial{\cal F}(\theta,\phi)}{\partial\theta} {\cal F}^{-1}(\theta,\phi) \widehat{S}\right)\;,\nonumber\\
\left.  \frac{\partial\epsilon}{\partial\phi^\prime}\right|_{\substack{\theta^\prime=\theta,\\ \phi^\prime=\phi}} &=&
\widehat{S}\cdot \left(\frac{\partial{\cal F}(\theta,\phi)}{\partial\phi} {\cal F}^{-1}(\theta,\phi) \widehat{S}\right) \;.
\end{eqnarray}
%\end{widetext}
The norm of the gradient
\begin{eqnarray}
{\cal G}(\theta,\phi;  \widehat{S}) & = & \left[ 
\left(\widehat{S}\cdot \left({\cal F}_\phi {\cal F}^{-1} \widehat{S}\right)\right)^2\right. +\nonumber \\
&& \left. \left(\widehat{S}\cdot \left({\cal F}_\theta {\cal F}^{-1} \widehat{S}\right)\right)^2 \right]^{1/2}\;,
\label{normsnrgrad}
\end{eqnarray}
${\cal F}_\theta = \partial {\cal F}/\partial\theta$ and ${\cal F}_\phi = \partial {\cal F}/\partial\phi$, 
is a convenient measure of the variability of the SNR  of a source as a function of its location. 

A high value of ${\cal G}(\theta,\phi;  \widehat{S}) $  implies that the source would have to occur at very special positions on the sky to be detectable. This is because  if it has an SNR above some detection threshold at one location, it may rapidly fall below it if it were to occur at a slightly different location. Consider two classes of sources, class A with a large associated ${\cal G}(\theta,\phi;  \widehat{S})$ and class B with low values of the same. Assume that events from both classes occur with the same comoving rate and that they have the same distribution on the sky. Then, due to the rapid fluctuation of the SNR for source class A, the area of the sky over which its members will be detectable would be significantly less than that of class B. It follows that the observed rate of the class A sources would be much reduced in comparison to class B. Therefore, even though the two sources may be equivalent in terms of, say,  the energy emitted in GWs, the high SNR variability source would be  detectable less often. Since a larger volume of space would 
required to detect such sources, it will naturally be harder to detect at a given false alarm rate.

We thus arrive at the main point of this section. Given a pair of misaligned detectors, we should include only those responses ${\bf S}$ in our search for the maximum of the LR for which ${\cal G}(\theta,\phi;\widehat{S})$ is low in some well defined sense. In this way, the detectability of sources with an intrinsically low observable rate will be reduced but that of more frequently observable sources will be enhanced. It is important to note that the above argument does not require any prior assumptions about the astrophysics of sources.  
Also note that $\cal G$ depends only on the orientation of ${\bf S}$ and not its magnitude. That is, it is only concerned with the shape of the  
waveforms of $h_+$ and $h_\times$ and not the distance to the source.

\begin{figure*}
\includegraphics[scale=0.7]{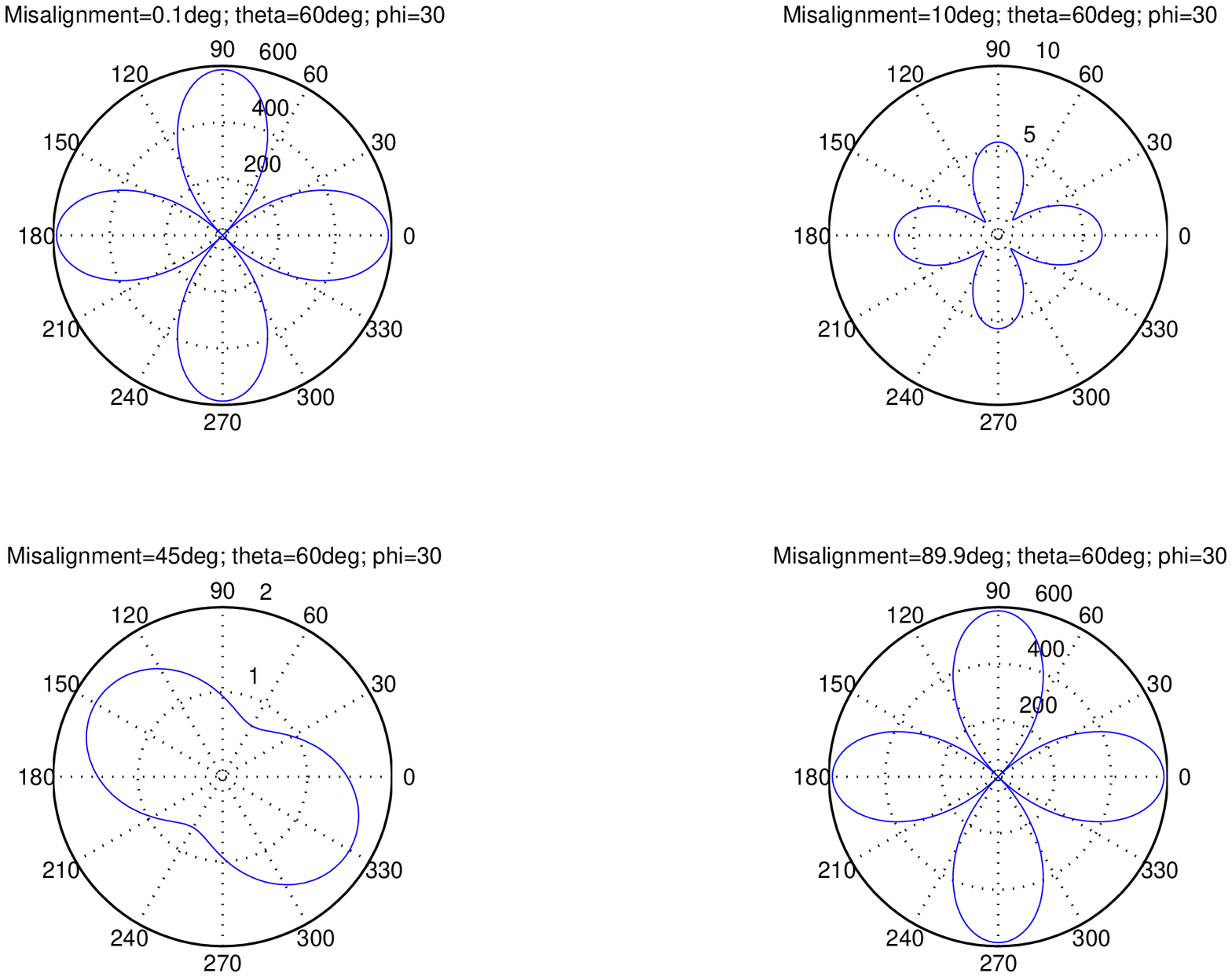}
\caption{\label{misalign_1}
Plots of ${\cal G}$ (Eq.~\ref{normsnrgrad}) for different misalignment angles. }
\end{figure*}

Figure~\ref{misalign_1} shows polar plots of ${\cal G}(\theta,\phi;  \widehat{S})$ as a function of $\widehat{S}$
 for fixed $\theta$, $\phi$ and different detector alignments. 
It is assumed that the detectors are colocated at the origin of the fiducial frame and coplanar with angles between their respective X arms of $\alpha$. The antenna patterns are, in this case,
\begin{equation}
{\cal F} = \left(\! \begin{array}{lr}
  \frac{1}{2} (1 + \cos^2 \theta)\cos 2\phi  &  \cos \theta \,\sin 2\phi \\\\
\frac{1}{2}(1 + \cos^2 \theta) \cos 2(\phi+\alpha)  &  \cos \theta \, \sin 2(\phi+\alpha)
\end{array}\!\right)\,.
\end{equation}
Each plot is over all orientations of the detector response vector $\widehat{S}$. First, we see that for almost coaligned detectors, $\alpha=0.1^\circ$,  If we limit the search for the maximum of $\lambda$ (c.f., Eq.~\ref{llrsummand}) over $\widehat{S}$ such that ${\cal G}(\theta,\phi;  \widehat{S}) \leq 10$, say,
then only detector responses with orientation very close to the $45^\circ$ and $-45^\circ$ line  will be allowed. This means that the instantaneous response in the two detectors should be nearly equal. From this example, we see that waveforms with low SNR variability are also the ones that lead to nearly identical responses in the two detectors and, as discussed above, are the ones which have relatively higher chances of being detected. This bears out our physical intuition that sources that produce very different responses in nearly aligned detectors should be rare. As $\alpha$ is increased, one sees that not only does ${\cal G}$ decrease but that it also approaches the same value for all orientation and, hence, all orientations of $\widehat{S}$ are now allowed. The pattern opens up fully when the detectors are at $45^\circ$ to each other.  As $\alpha$ approaches $90^\circ$, the responses are again
increasingly restricted around the $\pm 45^\circ$ lines.  

\begin{figure*}
\includegraphics[scale=0.7]{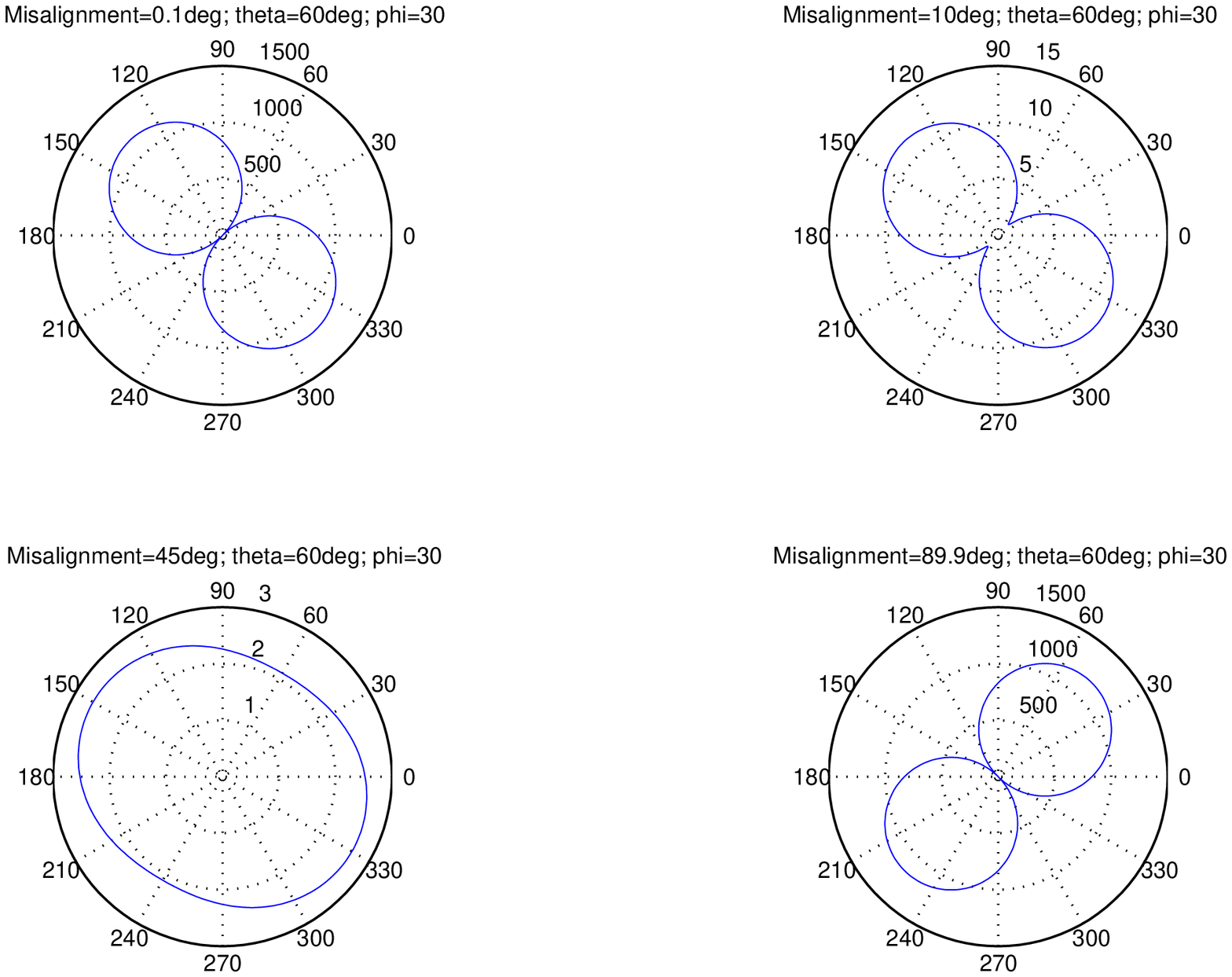}
\caption{\label{misalign_2}
Plots of $\Gamma$ (Eq.~\ref{snrgrad_upper}) for different misalignment angles. }
\end{figure*}

Though Fig.~\ref{misalign_1}  shows that SNR variability is a natural measure for restricting the orientation of $\widehat{S}$, the particular measure ${\cal G}(\theta,\phi;\widehat{S})$ allows both $\pm 45^\circ$ lines as the locii of restriction. For nearly aligned detectors, however, we expect detector responses to be  in-phase which means that the responses should lie in the vicinity of the $+45^\circ$ line. We introduce another function, $\Gamma(\theta,\phi;\widehat{S})$, that is also a measure of SNR variability but has better properties in this respect.
\begin{eqnarray}
\Gamma^2(\theta,\phi; \widehat{S}) & = &\left\|{\cal F}_\theta
{\cal F}^{-1} \widehat{S}\right\|^2+ \left\|{\cal F}_\phi
{\cal F}^{-1} \widehat{S}\right\|^2\;,
\label{snrgrad_upper}
\end{eqnarray}
In fact, from the Cauchy Schwartz inequality, it follows that
\begin{equation}
{\cal G}(\theta,\phi; \widehat{S}) \leq  \Gamma(\theta,\phi; \widehat{S}) \;.
\end{equation}
Hence, $\Gamma(\theta,\phi;\widehat{S})$ is an upper bound on 
SNR variability.
Figure~\ref{misalign_2} shows polar plots of $\Gamma(\theta,\phi; \widehat{S})$ for fixed $\theta$, $\phi$ and different misalignment 
angles $\alpha$. Now, for small misalignement ($\alpha=0.1^\circ$), one sees that only responses along the $+45^\circ$ line are allowed while for 
misalignement close to $90^\circ$, only responses that are close to $-45^\circ$ are allowed.  It makes sense, therefore, to use $\Gamma(\theta,\phi;\widehat{S})$ as a measure of SNR variability instead of ${\cal G}(\theta,\phi;\widehat{S})$.

In the above discussion, we considered the detection of a signal at a single instant using a matched filter, leading to the detection statistic ${\bf X}[k]\cdot{\bf S}[k]$. Burst signals can last over several sampling intervals and, in general, the detection statistic would in fact
 be a sum $\sum {\bf X}[k]\cdot{\bf S}[k]$ over a range of $k$. In this case, a high SNR variability response ${\bf S}[k]$ can be balanced by a very low variability response ${\bf S}[k^\prime]$ at some other instant, thus leading to a low variability for the overall SNR $ = [\sum \|{\bf S}\|^2]^{1/2}$. This would allow a broader range of signals to be used for maximizing the overall likelihood $\Lambda$ (c.f., Eq.~\ref{loglratio}).  Our choice of using instantaneous SNR variability is simply a conservative one since signals with low instantaneous SNR variability will necessarily have low variability of their overall SNR. Moreover, as discussed earlier, physical intuition suggests that the responses in closely aligned detectors should be nearly identical at each instant of time and not just on the average.   
%%%%%%%%%%%%%%%%%%%%%%%%%%%%%%%%%%%%%%%%%
\subsection{General network of detectors}
We will now generalize the SNR variability measure $\Gamma(\theta,\phi; \widehat{S})$, introduced in Eq.~\ref{snrgrad_upper},
 to the case of an arbitrary network of detectors. The relative difference in the observed SNR of the same source between
 two positions $(\theta,\phi)$ and
 $(\theta^\prime,\phi^\prime)$ is,
\begin{equation}
\epsilon(\theta,\phi,\theta^\prime,\phi^\prime;\widehat{S}) = \| {\cal F}(\theta^\prime,\phi^\prime){\cal H}^{-1}(\theta,\phi)
{\cal F}^T(\theta,\phi)
\widehat{S}\|-1\;.
\end{equation}
 As before, for a small displacement of the source,
\begin{eqnarray}
\left. \frac{\partial\epsilon}{\partial\theta^\prime}\right|_{\substack{\theta^\prime=\theta,\\ \phi^\prime=\phi}}&=& \widehat{S}\cdot \left( 
{\cal F}_\theta {\cal H}^{-1} {\cal F}^T \widehat{S}\right)\, ,\\
\left. \frac{\partial\epsilon}{\partial\phi^\prime}\right|_{\substack{\theta^\prime=\theta, \\ \phi^\prime=\phi}}&=& \widehat{S}\cdot \left( 
{\cal F}_\phi{\cal H}^{-1} {\cal F}^T\widehat{S}\right)\, ,
\end{eqnarray}
and our measure of SNR variability will be defined as,
\begin{eqnarray}
\Gamma^2(\theta,\phi;\widehat{S})  & = &  \left\|
{\cal F}_\theta {\cal H}^{-1}{\cal F}^T \widehat{S}\right\|^2+ \left\|{\cal F}_\phi {\cal H}^{-1} {\cal F}^T \widehat{S}\right\|^2.
\label{snrvardef_gen}
\end{eqnarray}
Here, $\widehat{S}$ is a unit vector in the $R^{N_d}$ space but lying in the response plane.

 For algebraic simplicity, it is convenient to express the response vector ${\bf S}$ in terms of a basis in the response plane rather than the full $R^{N_d}$
space. In order to do so, we need to fix two orthonormal vectors $\widehat{q}_0$ and $\widehat{q}_1$ on this plane. If $\widehat{e}_0$ and
$\widehat{e}_1$ are the two orthonormal eigenvectors of ${\cal H}$, corresponding to eigenvalues $\nu_0$ and $\nu_1$
respectively, then it is easily seen that for,
\begin{eqnarray}
\widehat{q}_0 &=& \frac{1}{\sqrt{|\nu_0|}}{\cal F}(\theta,\phi) \widehat{e}_0\\
\widehat{q}_1 &=& \frac{1}{\sqrt{|\nu_1|}}{\cal F}(\theta,\phi)\widehat{e}_1\\
\widehat{q}_0\cdot \widehat{q}_1&=& \widehat{e}_0^T {\cal H} \widehat{e}_1 = 0\;.\\
\|\widehat{q}_0\|&=&\|\widehat{q}_1\|=1\;.
\end{eqnarray}
Since, $\widehat{q}_0$ and $\widehat{q}_1$ have the same form as the detector responses (c.f. Eq.~\ref{ndetresponse}),
they are confined to the response plane.   Let $\gamma$ be the angle between $\widehat{\bf S}$ and $\widehat{q}_0$.
Then,
\begin{eqnarray}
{\bf S}&=&\|{\bf S}\| \left(\widehat{q}_0\cos\gamma+\widehat{q}_1\sin\gamma\right)\;,
\label{lincomb2d}
%&=& {\cal F}(\theta,\phi) \widetilde{S}(\gamma)\;,\nonumber\\
%\widetilde{S} &=&\frac{1}{\sqrt{|\nu_0|}} \widehat{e}_0 \cos\gamma+\frac{1}{\sqrt{|\nu_1|}} \widehat{e}_1 \sin\gamma\;.
%\label{widetildeS}
\end{eqnarray}
%By subsituting the above in Eq.~\ref{snrvardef_gen}, we get,
%\begin{eqnarray}
%\Gamma^2(\theta,\phi;\widehat{S}) & = & \Gamma^2(\theta,\phi;\gamma)\nonumber \\
%&=& \left\|{\cal F}_\theta \widetilde{S}(\gamma)\right\|^2 +\left\|{\cal F}_\phi \widetilde{S}(\gamma)\right\|^2
%\;,
%\label{gradsnr_final}
%\end{eqnarray}
 The SNR variability measure $\Gamma(\theta,\phi;\widehat{S})$ then becomes a function $\Gamma(\theta,\phi;\gamma)$  of $\gamma$. 

Recall Eq.~\ref{llrsummand} for the definition of the quantity $\lambda$ that
we want to maximize over the response vector ${\bf S}$. 
%Now, apply the constraint 
% that $\bf S$ should be confined to the response plane. This is implemented 
%by defining ${\bf S}$ following Eq.~\ref{lincomb2d},
%\begin{equation}
%{\bf S}=\rho (\widehat{q}_0 \cos\gamma + \widehat{q}_1\sin\gamma)\;.
%\end{equation}
%where $\rho = ||{\bf S}||$.
We can now rewrite it as, 
\begin{equation}
\lambda=\left( {\bf X}\cdot\widehat{q}_0\right)\rho \cos\gamma + \left({\bf X}\cdot
\widehat{q}_1\right)\rho  \sin\gamma - \frac{1}{2}\rho^2\;,
\label{lambda_mu0mu1}
\end{equation}
where $\rho = \| {\bf S}\|$.
From the above it is obvious that the solution for ${\bf S}$ that maximizes $\lambda$ is the projection of 
${\bf X}$ onto the response plane.

%%%%%%%%%%%%%%%%%%%%%%%%%%
\section{Penalized maximization of likelihood Ratio}
\label{penlike}
We have shown that SNR variability, as measured by $\Gamma(\theta,\phi;\gamma)$ defined in Eq.~\ref{snrvardef_gen}, correlates
well with how improbable a solution of the LR maximization problem is. Improbable waveforms $h_+[k]$ and $h_\times[k]$ that lead to very
different responses in nearly aligned detectors also correspond to a fast variation in the detectability of the source as it is displaced on
the sky. Such sources naturally become rare compared to probable ones. In this Section, we construct a 
detection statistic that incorporates $\Gamma(\theta,\phi;\gamma)$ to
restrict the influence of improbable solutions on the maximization of the LR. 

\subsection{Algebraic simplifications}
We begin by making some algebraic simplifications. 
Substituting Eq.~\ref{lincomb2d} into~\ref{snrvardef_gen}, we get,
%\begin{eqnarray}
%\Gamma^2(\theta,\phi;\gamma) &=& \frac{\cos^2\gamma}{|\nu_0|}\left( \left\| {\cal F}_\theta \widehat{e}_0\right\|^2 + 
%\left\| {\cal F}_\phi \widehat{e}_0\right\|^2
%\right) +\nonumber\\
%&& \frac{\sin^2\gamma}{|\nu_1|} \left( \left\| {\cal F}_\theta \widehat{e}_1\right\|^2 + 
%\left\| {\cal F}_\phi \widehat{e}_1\right\|^2 \right)+\nonumber\\
%&& \frac{2}{\sqrt{|\nu_0\nu_1|}} \left[
%\left( {\cal F}_\theta\widehat{e}_0
%\right).\left(
%{\cal F}_\theta \widehat{e}_1
%\right) + \right.\nonumber\\
%&&
%\left. \left( {\cal F}_\phi \widehat{e}_0
%\right).\left(
%{\cal F}_\phi \widehat{e}_1
%\right)
%\right]\cos\gamma\sin\gamma
%\label{gammasq}
%\end{eqnarray}
%We can rewrite Eq.~\ref{gammasq}
%as,
%%Put equations on separate lines
\begin{eqnarray}
\Gamma^2(\theta,\phi,\gamma) &=& \left(\begin{array}{cc} \cos\gamma & \sin\gamma \end{array}\right)
{\cal M}
\left(
\begin{array}{c} \cos\gamma \\ \sin\gamma \end{array}\right) \;,\\
{\cal M} &=&  \left(\begin{array}{cc} A & C \\ C & B\end{array}\right)\;,\\
A &= & \frac{1}{|\nu_0|}\left(
\|{\cal F}_\theta \widehat{e}_0\|^2 + 
\|{\cal F}_\phi \widehat{e}_0\|^2
\right)\;,\\
B  &=& \frac{1}{|\nu_1|}
\left(
\|{\cal F}_\theta\widehat{e}_1\|^2 + 
\|{\cal F}_\phi \widehat{e}_1\|^2
\right)\;,\\
C &=& \frac{1}{\sqrt{|\nu_0\nu_1|}}\left[
\left( {\cal F}_\theta \widehat{e}_0
\right)\cdot \left( {\cal F}_\theta \widehat{e}_1
\right) + \right.\nonumber \\
&&\left.
\left( {\cal F}_\phi \widehat{e}_0
\right)\cdot\left( {\cal F}_\phi \widehat{e}_1
\right)\right]\;.
\end{eqnarray}
We now perform another change of basis in the response plane.
Let the eigenvalues and eigenvectors of ${\cal M}$ be $\mu_0$, $\mu_1\leq \mu_0$ and $\widehat{m}_0$ and $\widehat{m}_1$ respectively. Since $\widehat{m}_0$ and $\widehat{m}_1$ form an orthonormal basis,
\begin{equation}
\left(
\begin{array}{c}
\cos\gamma \\ \sin\gamma
\end{array} \right) =  \widehat{m}_0 \cos\psi+ \widehat{m}_1\sin\psi \;,
\label{cossingamma}
\end{equation}
for some angle $\psi$ between the response vector ${\bf S}$ and $\widehat{m}_0$.
Then, 
\begin{equation}
\Gamma^2(\theta,\phi;\gamma)  = \mu_0 \cos^2\psi + \mu_1 \sin^2\psi\;.
\label{gamma_psi}
\end{equation}
Henceforth, we denote $\Gamma^2(\theta,\phi;\gamma)$ as $\Gamma^2(\theta,\phi;\psi)$.

Using the orthonormal basis vectors $\widehat{m}_0$ and $\widehat{m}_1$, we can state
\begin{equation}
\left(\begin{array}{c}
{\bf X}\cdot \widehat{q}_0\\
{\bf X}\cdot \widehat{q}_1 
\end{array}\right) = \alpha_x \widehat{m}_0 +\beta_x \widehat{m}_1\;,
\label{cossinx}
\end{equation} 
where $\alpha_x$ and $\beta_x$ denote the components along $\widehat{m}_0$ and $\widehat{m}_1$ respectively.
From Eq.~\ref{lambda_mu0mu1} and~\ref{cossinx}, it follows that,
\begin{eqnarray}
\lambda &=& \rho \left(\begin{array}{cc} \cos\gamma & \sin\gamma \end{array}\right) \left(\begin{array}{c} {\bf X}\cdot \widehat{q}_0\\
{\bf X}\cdot \widehat{q}_1 \end{array}\right) -\frac{1}{2}\rho^2 \nonumber \;,\\
&=& \rho\left(\alpha_x\cos\psi + \beta_x \sin\psi
\right) - \frac{1}{2}\rho^2 \;,\nonumber\\
&=& \rho l_x \cos\left(\psi -\phi_x\right) -\frac{1}{2}\rho^2\;,\\
l_x & =& \sqrt{\alpha_x^2 +\beta_x^2}\;,\\
\phi_x &=& \tan^{-1}\left(\frac{\beta_x}{\alpha_x}\right)\;.
\end{eqnarray}
The  maximization of $\lambda$ has to be performed over $\rho$ and $\psi$.  Thus, the SL solution is $\rho=l_x$, the length of the projection of 
${\bf X}$ onto the response plane, and $\psi=\phi_x$ which is the angle between the projection and the $\widehat{m}_0$ vector. %%%%%%%%%%%%%%%%%%%%%%%%%%%%%%%%%%%
\subsection{Penalized maximization}
Now we introduce one possible way to incorporate $\Gamma(\theta,\phi,\psi)$ into the maximization of $\lambda$.
We call this method {\em penalized maximum likelihood ratio} (PMLR).
We propose to maximize $\lambda^\prime$,
\begin{equation}
\lambda^\prime = \lambda - \alpha \Gamma^2(\theta,\phi,\psi)\;,
\label{penalizedlr}
\end{equation}
where $\alpha \geq 0$ is a parameter under our control. The extra term
$\alpha \Gamma^2(\theta,\phi,\psi)$ acts as a penalty that prevents the solution from readily converging to the data 
vector $\bf X$. This resolves the two detector paradox (c.f., Section~\ref{2dp}). 

Once the solution $\widehat{\rho}$, 
$\widehat{\psi}$ that maximizes 
$\lambda^\prime$ in Eq.~\ref{penalizedlr}
is found, we find the value of $\lambda$ at this solution (not $\lambda^\prime$).
%Recall that $\lambda$ was the logarithm of the likelihood ratio $\Lambda$, (c.f. Eq.~\ref{loglratio})
% at a single time instant. 
Our final data functional will be
the sum of the values of $\lambda$, found as above, over all samples of the data ${\bf X}$. Reintroducing the
time index and denoting the resulting functional as $\Lambda_{\rm PL}(\theta,\phi)$, we get
\begin{widetext}
\begin{equation}
\Lambda_{\rm PL}(\theta,\phi)  = \sum_j \max_{\rho,\psi}\left[
 \rho l_x[j] \cos\left(\psi[j] -\phi_x[j]\right) -\frac{1}{2}\rho[j]^2 - \alpha (\mu_0 \cos^2\psi[j]+\mu_1\sin^2\psi[j])\right]\;.
\end{equation}
\end{widetext}
The global maximum of $\Lambda_{\rm PL}(\theta,\phi)$, over $\theta$ and $\phi$, will serve as the
 final detection statistic.

The solution $\widehat{\rho}$,  $\widehat{\psi}$ that maximizes $\lambda^\prime$ is easily found.
\begin{eqnarray}
\widehat{\rho} &=& l_x \cos(\widehat{\psi}-\phi_x)\;,\label{rho_solution}\\
\tan( 2\widehat{\psi}) & =& \frac{\sin(2 \phi_x)}{\cos (2 \phi_x) - 2 \alpha (\mu_0 -\mu_1)/l_x^2}\;.
\label{psi_solution}
\end{eqnarray}
Note that when $\alpha=0$, we recover the standard likelihood ratio statistic $\Lambda_{SL}(\theta,\phi)$
wherein the solution for $\bf S$ is the data $\bf X$ projected on
the 2D response plane. Also, the importance of the penalty term diminishes as $l_x^{-2}$. If a signal is present, the 
latter increases with an increase in the instantaneous signal amplitude. Hence, whenever the signal amplitude is strong enough, the effect of
the penalty term is reduced at that instant and the wave form estimators $\widehat{\psi}$ and $\widehat{\rho}$ converge to
the SL estimates. 

\begin{figure*}
\includegraphics[scale=0.4]{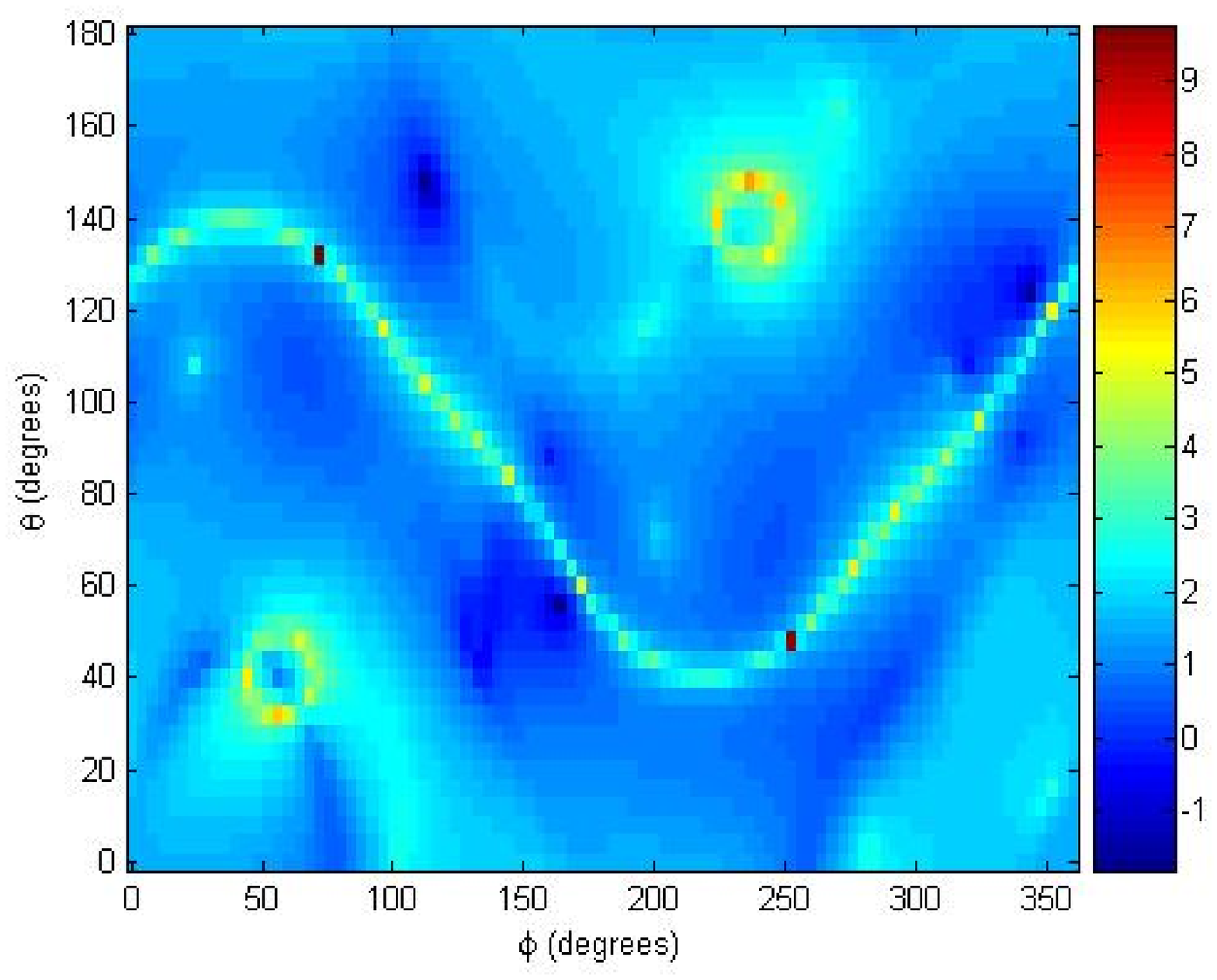}
\includegraphics[scale=0.4]{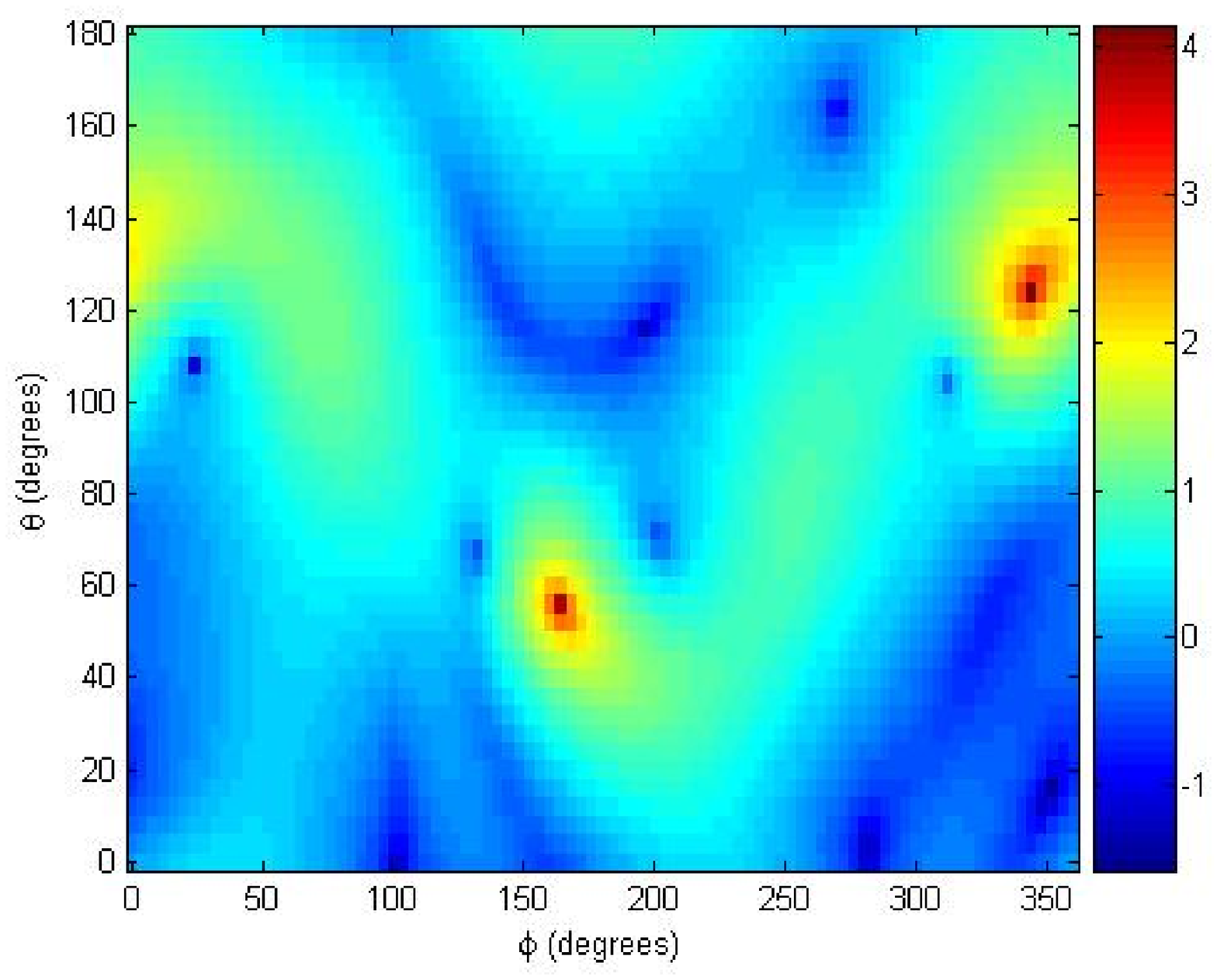}\\
\includegraphics[scale=0.4]{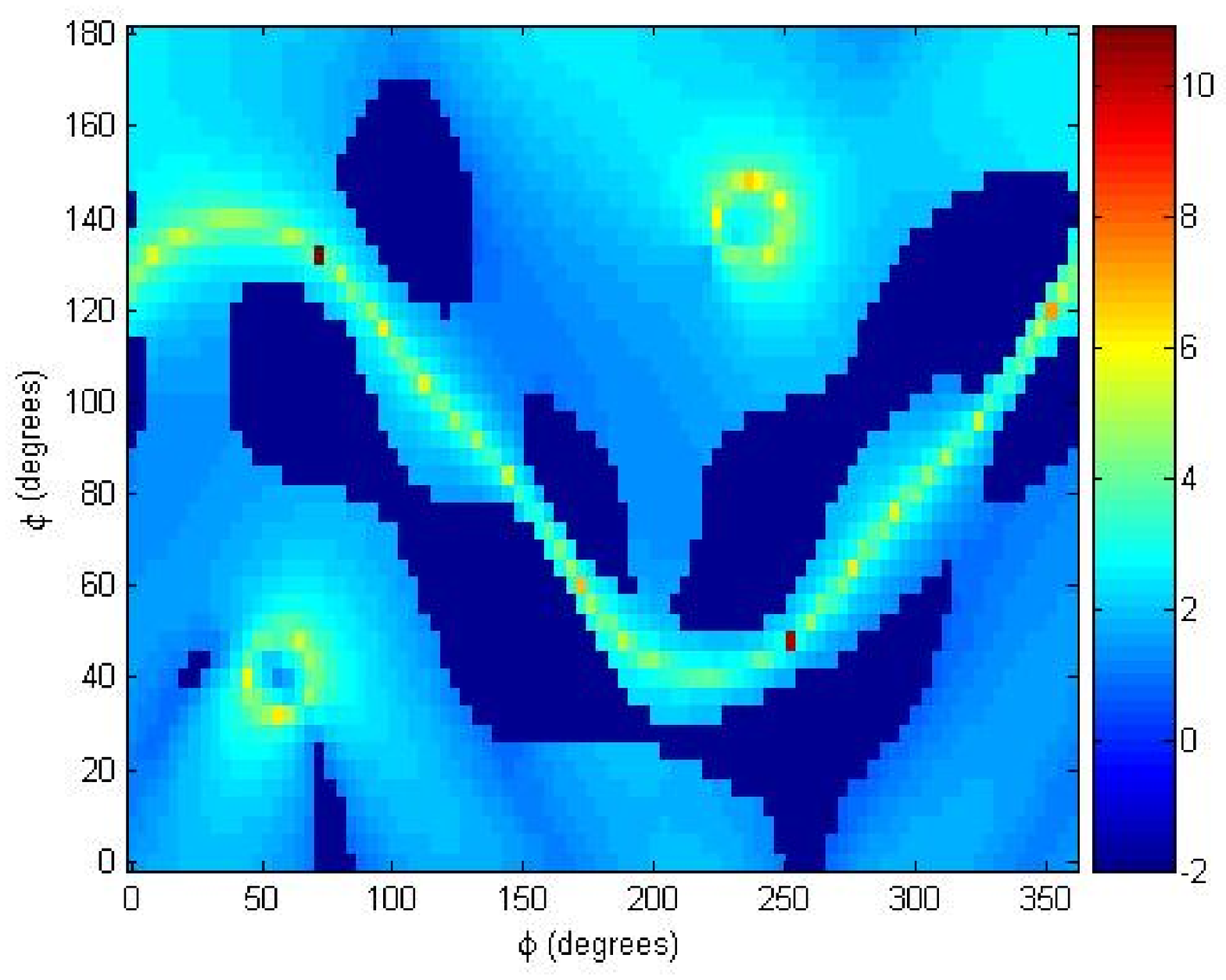}
\includegraphics[scale=0.4]{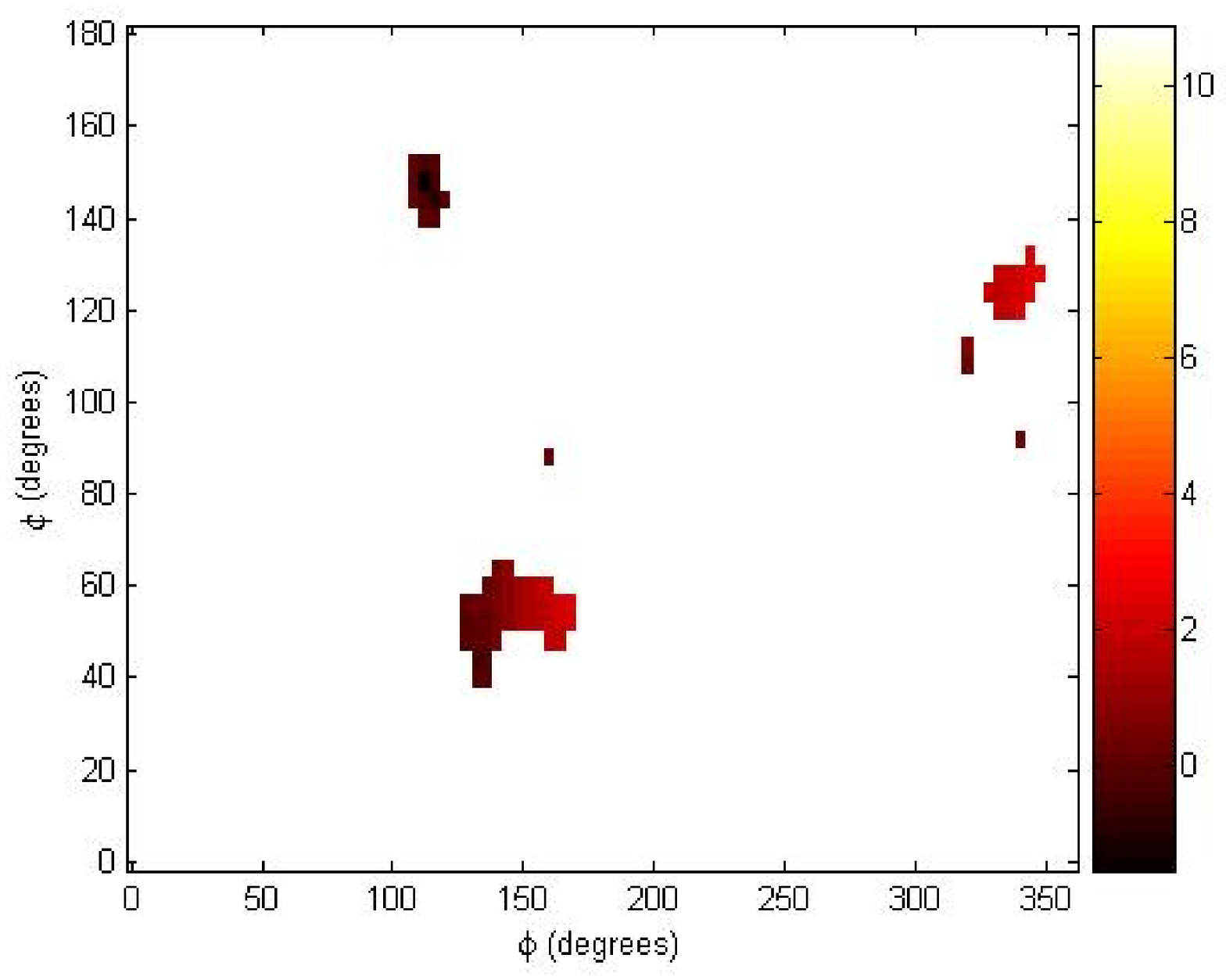}
\caption{The difference in eigenvalues $\mu_0$ and $\mu_1$ plotted as a function of sky-position angles $\theta$ and $\phi$. 
(All panels display the logarithm to base 10 of $\mu_0-\mu_1$.)
The top left panel is for a network consisting of three LIGO detectors: two at Hanford and one at Livingston. The
top right panel is for a network consisting of one LIGO detector at Livingston, one GEO and one VIRGO detector.
 The first network (top left), which is more aligned, shows much larger values of $\mu_0-\mu_1$ than
the second (top right) which has more misalignement between the detectors.
In the bottom left panel, the part of the sky where 
$\mu_0-\mu_1$ for the first network is $\leq 10$~times that for the second network  has been suppressed. It is clear
that this is a smaller fraction of the sky compared to the part where $\mu_0-\mu_1$ for the first network is $\geq 10$~times that of the 
second network.
In the bottom right panel, only the part of the sky where  $\mu_0-\mu_1$ for the second network is greater than or equal to that 
for the first network is shown. \label{eval_diff}}
\end{figure*}

The term $\mu_0-\mu_1$, which is always positive by definition, is a
measure of how misaligned the detectors in a network are. This difference is reduced with an increase in misalignment, hence reducing the effect of
the penalty term. %Note that $l_x$, $\phi_x$, $\mu_0$ and $\mu_1$ are all dependent on the sky position $\theta$, $\phi$.   
Fig.~\ref{eval_diff} 
shows $\mu_0 -\mu_1$ as a function of sky position for two networks: (1) three LIGO detectors:
one at Livingston, and two at Hanford, and (2)
one LIGO detector at Livingston,  one GEO600 and one VIRGO detector. 
Fig~\ref{eval_diff} also shows the ratio of the differences in eigenvalues for the two networks. It is clear that the difference in eigenvalues for the first
network, which consists of nearly aligned detectors, is larger than that for the second network over a large fraction of the sky. 

The statistic $\Lambda_{\rm PL}(\theta,\phi)$ solves the two detector paradox by continuously bridging the aligned and misaligned detectors cases.
From Eq.~\ref{psi_solution}, it is clear that  as $\mu_0-\mu_1$ increases, i.e., as the detectors become more aligned,
the denominator 
increases in the  negative direction while the numerator, $\sin 2 \phi_x$, stays the same. Hence, 
$2 \psi \rightarrow \pi$ which implies $\psi \rightarrow \pi/2$. From Eq.~\ref{gamma_psi}, the penalty function $\Gamma^2$
has its lowest value for $\psi=\pi/2$. This is the direction in the response plane along which
 all the responses in a network get oriented
as the detectors are aligned. Therefore, with progressive alignment of the detectors in a network,
 the solution $\widehat{\psi}$ is forced towards equal responses in all detectors, thus leading smoothly into the 
completely aligned case. In contrast, 
for the standard likelihood statistic $\Lambda_{\rm SL}(\theta,\phi)$ ($=\Lambda_{\rm PL}$ when
$\alpha=0$), there exists a fundamental discontinuity 
between the solutions for aligned and misaligned cases: 
For the aligned case $\widehat{\psi}=\pi/2$, while $\widehat{\psi}=\phi_x$ for 
the misaligned case.

Finally, the penalty term, $\Gamma^2$, depends only the direction of the instantaneous network response vector ${\bf S}$. 
It does not restrict its length. Hence, the magnitude of ${\bf S}$ at two distinct time instants can be arbitrarily different. Thus,
even white noise response waveforms are allowed as solutions. The only requirement is on the mutual
 consistency of detector responses at each instant of time.

%Finally, we point out a significant difference between the penalized and the constraint likelihood method. 
% The hard constraint statistic is obtained from one component of the likelihood functional
% and it remains so irrespective of the actual data (see Sec.~\ref{2dp}). The soft constraint statistic involves a weighted sum
%of the two components of the likelihood functional but the weight factor does not depend on the data. Thus, the statistics
%are never equal to standard likelihood. 
% In the penalized maximization approach, on the other hand, $\lambda^\prime$ can 
%converge to the standard likelihood solution for sufficiently large $l_x$. The penalty factor $\alpha$
%controls the rate of the convergence as a function of signal amplitude. 

%%%%%%%%%%%%%%%%
\subsection{Connection with Bayesian analysis}
The penalized likelihood approach, Eq.~\ref{penalizedlr}, has a straightforward
 connection to Bayesian~\cite{bayes} analysis. In the latter approach, the posterior degree of belief over the space of waveforms, or equivalently $\rho$ and $\psi$ (c.f., Eq.~ref{}), is obtained using Bayes' law,
\begin{equation}
p(\rho,\psi|{\bf X}) = \frac{p(\rho,\psi)p({\bf X}|\rho,\psi)}{p({\bf X})}\;,
\end{equation}
where $p({\bf X}|\rho,\psi)$ is simply the likelihood $\lambda$ in Eq.~\ref{llrsummand}. The maximum a posteriori 
 Bayesian estimates for the true $\rho$ and $\psi$ would be the ones for which $p(\rho,\psi |{\bf X})$ is maximized. This is equivalent to maximizing the logarithm of $p(\rho,\psi |{\bf X})$ and, hence, the sum of $\lambda$ and $\log p(\rho,\psi)$. Comparison with Eq.~\ref{penalizedlr} shows that $1/\Gamma^2$ acts as a prior. The difference is that this prior is not normalized in Eq.~\ref{penalizedlr} but is instead multiplied by $\alpha$. Of course, $\alpha$ can be chosen to be the integral of $1/\Gamma^2$ 
over $\psi$ thus reducing the penalized maximization to a true Bayesian analysis.  

%Maximum entropy estimation  

%%%%%%%%%%%%%%%%%%%%%%%%%%%%%%%%%%%%%%%%%%%%%%%
\section{Results}
\label{results}
The PMLR method was implemented in Matlab~\cite{matlab} and its performance was quantified with the 
Monte Carlo simulation code previously used in~\cite{Klimenko+etal:2005}. We give a brief recapitulation of
 the simulation scheme: Binary Black Hole merger waveforms from~\cite{lazarus} are injected into
short stretches of white noise, one waveform per stretch. The signals correspond to face-on binaries scattered uniformly on the sky
with randomly selected polarization angles. The signals have a variable matched filtering SNR 
due to their sky position and polarization, but an overall factor $G$ is used to scale the amplitudes of the two polarizations $h_+(t)$
and $h_\times (t)$. The sky-averaged SNR of the injected signals is approximately related to G as $SNR\simeq 2.3 G$.
%The detectors are considered to be identical and only their 
%orientations and geographical locations are used faithfully. 
Further details can be found in~\cite{Klimenko+etal:2005}.

\begin{figure*}[t]
\includegraphics[scale=0.7]{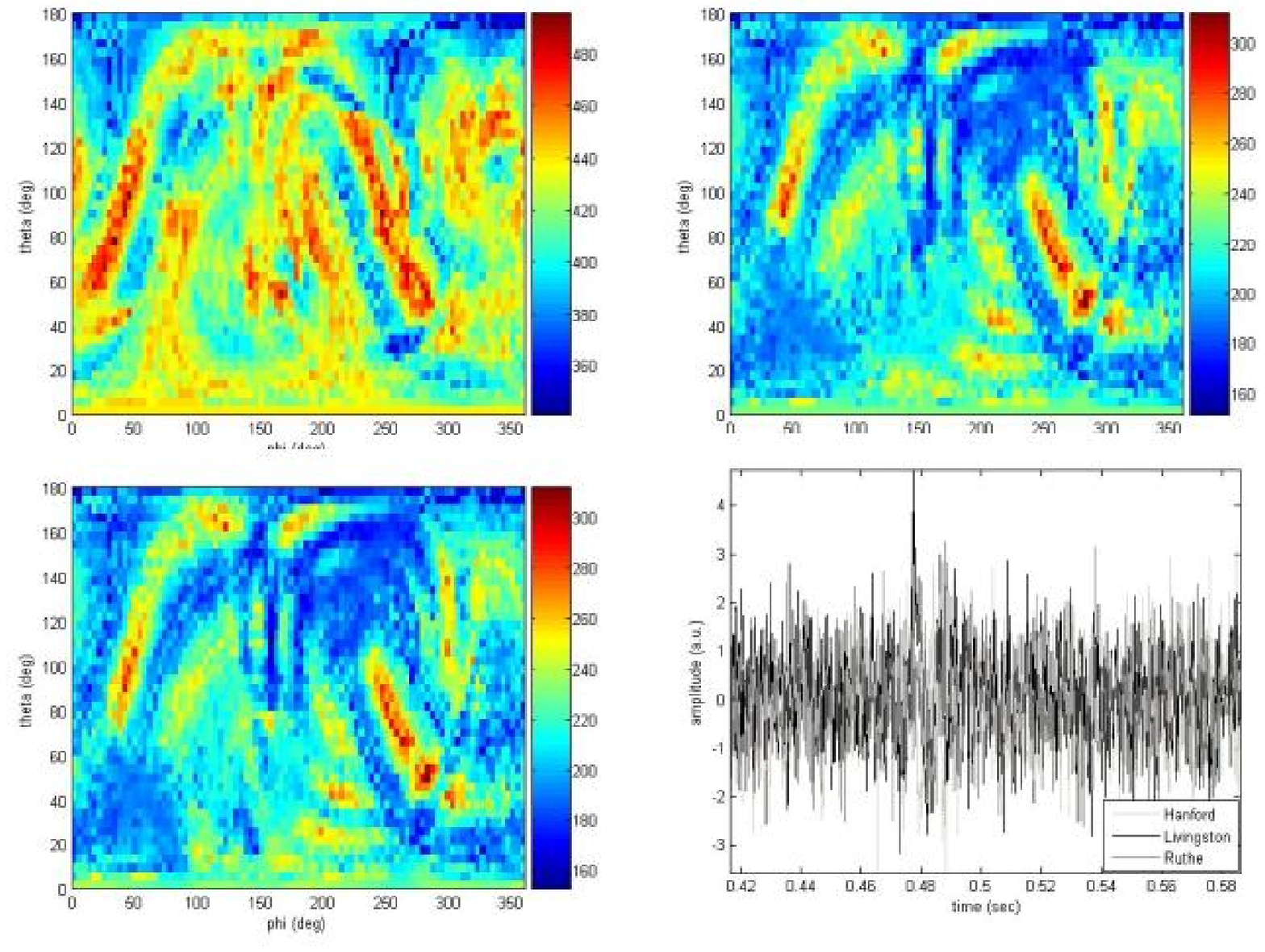}
\caption{\label{comp_1} Comparison of SL (top left), constraint likelihood (hard constraint) (top right) and PMLR (bottom left)
  for $\alpha=1500$. The source is at $\theta=50^\circ$, $\phi=280^\circ$ and has an 
amplitude scale factor of $G=3$. The network consists of one detector at Hanford, one at Livingston and GEO. 
The detector 
responses, assuming white Gaussian noise, are shown in the bottom right panel. }
\end{figure*}

Consider, first, some case studies to give a qualitative feel for the 
performance of PMLR compared to the SL and constraint likelihood methods.
Fig~\ref{comp_1} compares SL, constraint likelihood (hard constraint) and PMLR (with $\alpha=1500$) for the same signal 
using the three detector network: Hanford, Livingston and GEO. 
The true sky position of the source was $\theta=50^\circ$, $\phi=280^\circ$. The best estimates for sky
position (in degrees)
 were (1) $\theta=76^\circ$, $\phi=32^\circ$ for SL and (2) $\theta=48^\circ$, $\phi=280^\circ$
for PMLR and constraint likelihood. Note that, compared to SL,
 the sky maps for constraint likelihood and PMLR are less noisy and show more contrast between the 
region in which the signal is located and the rest of the sky. 

\begin{figure*}
\includegraphics[scale=0.8]{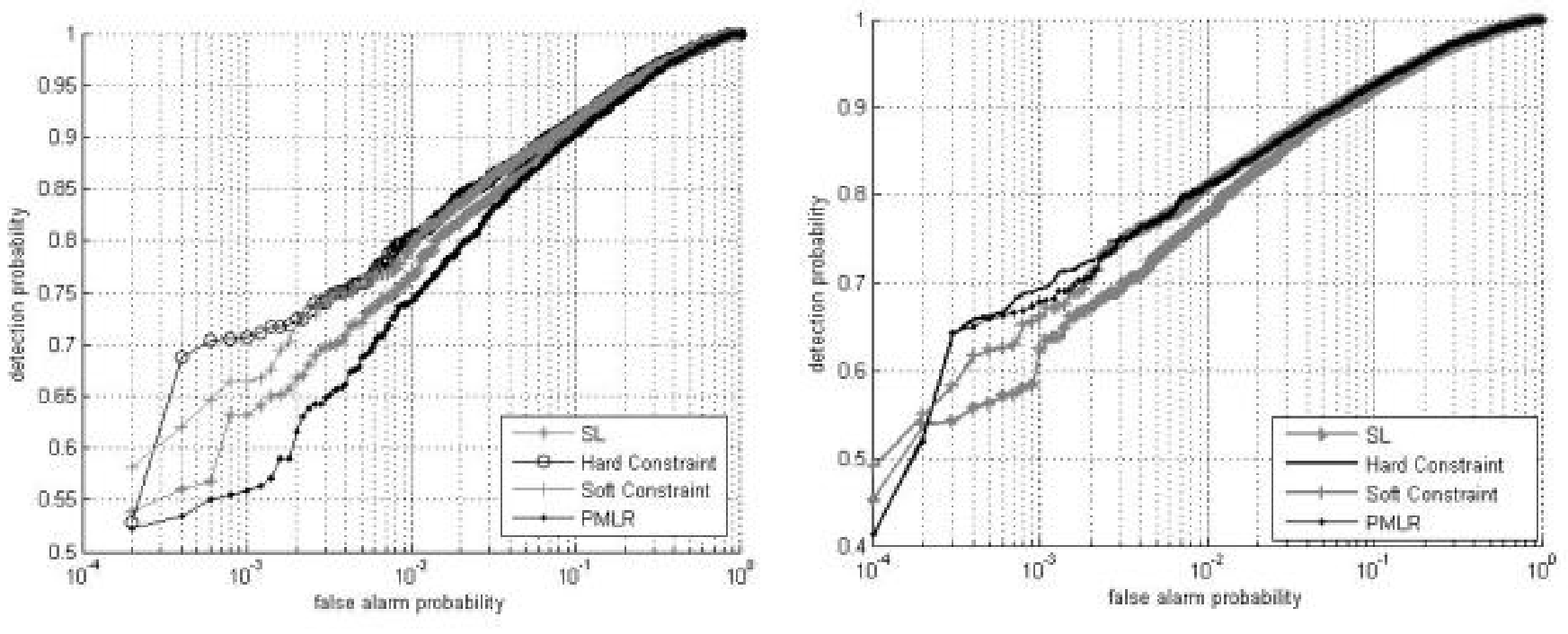}
\caption{Receiver Operating Characteristics for SL, constraint likelihood (both hard and soft constraint) and PMLR.
The value of the penality factor $\alpha=150$ and $\alpha=1500$ for the left andright panels respectively.
The number of trials was 5000 for both the left and right plots. The signal amplitude factor for this simulation was
$G=3.0$ which corresponds to an average SNR of $\simeq 7.0$. The network chosen was one detector
 at Hanford, one at Livingston and one
GEO detector.
\label{roc_curves}} 
\end{figure*}

The simulation code finds out the fraction of all injected binaries detected by the 
PMLR (or other) method. The detection statistic for all methods  is the maximum value over the respective 
skymap (e.g. Fig~\ref{comp_1}). Fig~\ref{roc_curves} shows
 the receiver operating characteristics (ROC) for SL, constraint likelihood (both hard and soft constraint)
and PMLR (with $\alpha=150$ and $1500$). The measured fraction of detected signals is plotted 
against the false alarm rate corresponding to the detection threshold used. It is clear from the figure
that PMLR can vary continuously in performance, ranging from SL for low values of $\alpha$
to the hard constraint method for high values.

\begin{figure}
\includegraphics[scale=0.6]{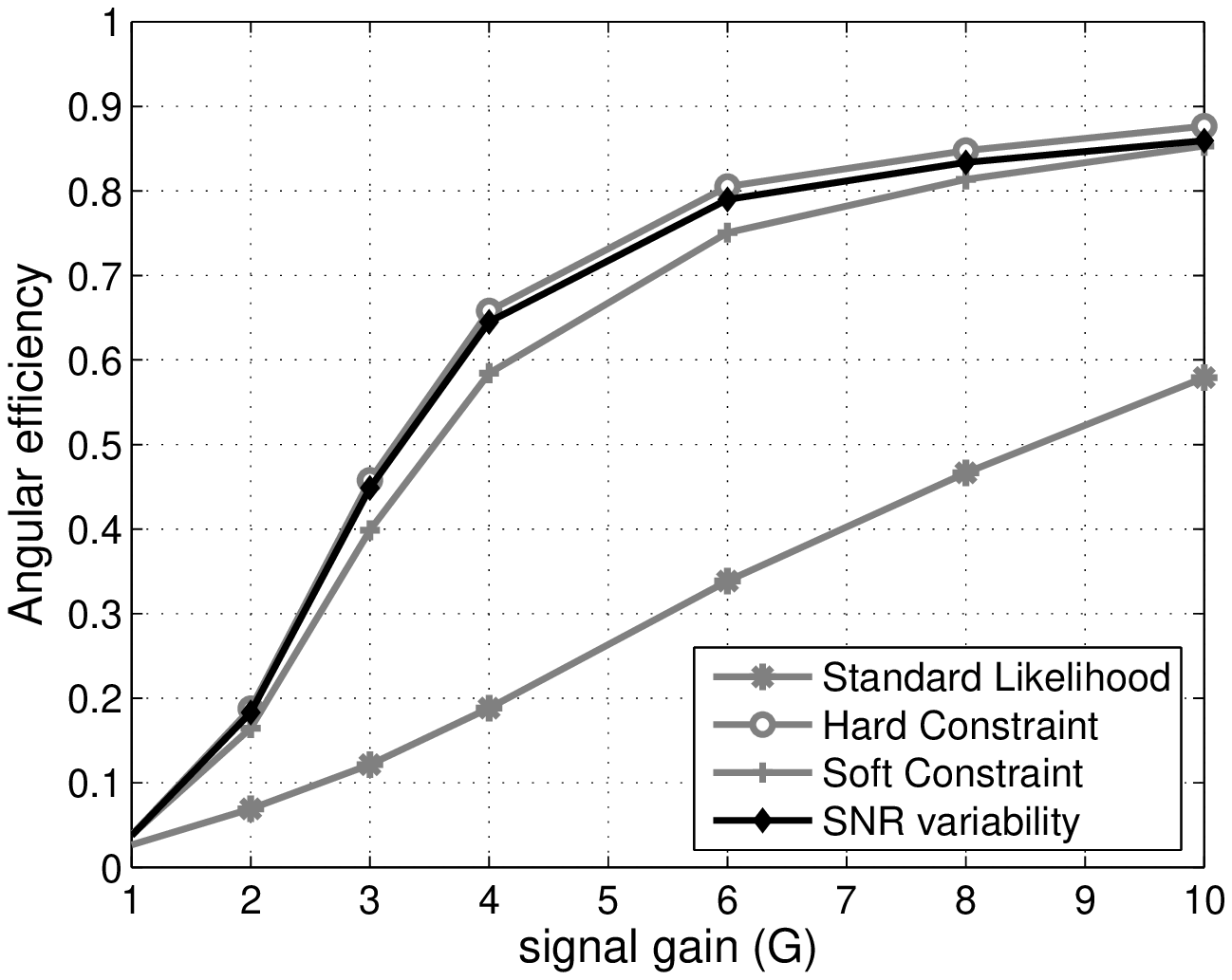}
\caption{\label{snrgrad_comp_angeff} 
The fraction of signals detected within $8^\circ$ of their true positions. As the signal SNR decreases, all methods lose both
detection efficiency and incur large errors in direction estimation. At the same time, as the SNR increases, all methods 
converge to the true signal location in almost all cases. The difference between the methods is, therefore, maximum for
intermediated SNRs.}
\end{figure}

The simulation code also estimates the error in sky position reconstruction. Fig.~\ref{snrgrad_comp_angeff} shows the 
fraction of the injected signals that were detected within 8 degrees of their true position, for a range
of the scale factor $G$. The PMLR method ($\alpha=1500$) has practically the same performance as
the hard constraint method and both perform somewhat better than the soft constraint method. The performance of
 SL, however, is significantly worse than all the other methods. This can also be seen qualitatively in the sky maps
of the various methods (Fig.~\ref{comp_1}) where the SL sky map always has less ``contrast" around the true signal location
compared to the constraint likelihood and PMLR methods.
%%%%%%%%%%%%%%%%%%%%%%%%%%%%%%%%%%%%%%%%%%%%%%%
\section{Conclusion}
There exists a serious problem, called the two detector paradox, in the formal
application of the standard likelihood method to the detection of
burst GW signals. We resolve this problem by imposing a restriction on the
detector responses allowed as candidate solutions for maximizing the likelihood functional. The restriction is 
imposed through a penalizing function that measures the variability of
the SNR of a candidate solution when the corresponding source is displaced slightly on the
sky.  High SNR variability sources lead to very different detector responses
even when the detectors are closely aligned. The standard likelihood method
allows such sources as perfectly valid candidate solutions. However, such abnormal sources have an inherently
reduced observability. The proposed method (PMLR) sacrifices the detectability of suuch sources while
boosting the sensitivity and sky position reconstruction accuracy for normal
signals. 

PMLR contains a free parameter, the penalty factor $\alpha$. For $\alpha=0$ PMLR
simply reduces to the SL method. However,
 method is not very sensitive to the exact value of $\alpha$, and the sky map of
the likelihood
converges rapidly as $\alpha$ is made large. The role of $\alpha$ needs to be 
investigated further. 

Results from numerical simulations show that the performance of the PMLR method 
can be tuned, using $\alpha$, to vary over the range covered by the standard likelihood  and 
the constraint
likelihood methods introduced in KMRM. For the type of source population used in the simulations, a high
value of $\alpha$ yields the same performance as the hard constraint method.  The ROC curves are
mildly better compared to standard likelihood but the accuracy of source direction reconstruction is 
significantly better.

PMLR is also related to the Bayesian approach. Work in the latter direction
has focussed on using priors that suppress candidate solutions that are
not smooth. (This is equivalent to the requirement that the signals have
a smaller frequency bandwidth than the noise.) However,
the penalty in PMLR is not based on the smoothness of signals in time. It is on signals that
lead to widely different responses in nearly aligned detectors. In particular, a 
white noise burst signal is perfectly admissible as a candidate solution in PMLR.
If smoothness constraints are desired, they can be added on to the PMLR restriction. 

The approach introduced in this paper can be extended to include 
variability of observed SNR as a function of source inclination to line of sight and the 
projected orientation of the source on the sky. This extension may result
in a penalizing function with better performance than the one used in this paper.

%%%%%%%%%%%%%%%%%%%%%%%%%%%%%%%%%%%%%%%%%%%%%%%
\begin{acknowledgements}
We thank an anonymous referee from the LIGO Science Collaboration for useful comments on the text.
This work was supported by the US National Science Foundation grants PHY-0244902, PHY-0070854 to the University of Florida, Gainesville and
NASA grant NAG5-13396 to the Center for Gravitational Wave Astronomy at the University of Texas at Brownsville. M.R.~was supported by NSF awards
PHYS 02-44902, PHYS 03-26281 and the NSF Center for Gravitational Wave Physics. The Center for Gravitational Wave Physics is funded by the
National Science Foundation under cooperative agreement PHY-01-14375.
\end{acknowledgements}
%%%%%%%%%%%%%%%%%%%%%%%%%%%%%%%%%%%%%%%%%%%%%%%
%\input{snrgrad_refs}

%%%%%%%%%%%%%%%%%%%%%%%%%%%%%%%%%%%%%%%%%%%%%%%

\begin{thebibliography}{99}
\bibitem{LIGO}A. Abramovici {\em et al},
 {\em Science} {\bf 256} 325-333 (1992).

\bibitem{VIRGO}F. Acernese {\em et al.},  Class. Quantum Grav. {\bf 21}, S385 (2004).

\bibitem{GEO600} B. Willke {\em et al.}, Class. Quantum Grav. {\bf 21}, S417 (2004).

\bibitem{TAMA}M.  Ando and the TAMA collaboration, Class. Quantum  Grav. {\bf 19}, 1409 (2002).

\bibitem{IGEC} IGEC: International Gravitational wave Event Collaboration, 
{\tt http://igec.lnl.infn.it/}

\bibitem{Stuart+Ord:v2} A.~Stuart, K.~Ord, {\em Kendall's Advanced Theory of Statistics}, Vol.2, $5^{\rm th}$ Edition (Edward Arnold, 1991).

\bibitem{finn} L.~S.~Finn, Phys. Rev. D {\bf 63}, 102001 (2001).

\bibitem{bose+etal}A.~Pai, S.~Dhurandhar, S.~Bose,  { Phys. Rev.} D {\bf 64}, 042004 (2001).

%\bibitem{stochastic}B.~Allen and J.D.~Romano, {\em Phys.\ Rev.\ D\,} {\bf 59}, 102001 (1999).

\bibitem{corecollapse}Kimberly C.B. New, ``Gravitational Waves from Gravitational Collapse", {\em Living Rev. Relativity} {\bf 6}, (2003), 2. URL (cited on June 16, 2005): http://www.livingreviews.org/lrr-2003-2.

\bibitem{Kobayashi+Meszaros}S.~Kobayashi, P.~M\'esz\'aros, { Astrophys. J.} {\bf 589}, 861 (2003).

\bibitem{flanagan+hughes:I}\'E.~Flanagan, S.~A.~Hughes, { Phys.~Rev.}~D {\bf 57}, 4535 (1998).

\bibitem{Flanagan+Hughes:II} \'E.~Flanagan, S.~A.~Hughes, { Phys.~Rev.}~D {\bf 57}, 4566 (1998).

\bibitem{mohanty+etal:gwdaw8} S.~Mohanty {\em et al}, { Class.~Quantum Grav.} {\bf  21}, S1831 (2004).

\bibitem{johnston}Wm.~R.~Johnston, M.S. thesis,
The University of Texas at Brownsville, 2004.

\bibitem{Klimenko+etal:2005} S.~Klimenko {\em et al}, Phys. Rev. D {\bf 72}, 122002 (2005).

\bibitem{guersel+tinto} Y.~G\"ursel, M.~Tinto, { Phys.~Rev.}~D {\bf 40}, 3884 (1989).

\bibitem{sylvestre}J.~Sylvestre, { Phys.~Rev.}~D {\bf 68}, 102005 (2003).

\bibitem{bayes} A.~O'Hagan, J.~Forster, {\em Kendall's Advanced Theory of Statistics, Vol.~2B: Bayesian Inference}, (Arnold, London, 2004).

\bibitem{Anderson+etal}W.~Anderson {\em et al}, {  Phys. Rev. D} {\bf 63}, 042003 (2001).

\bibitem{Brady+Mazumdar} P. R. Brady, S. Ray-Majumder, Class. Quantum Grav. {\bf 21}, S1839 (2004). 

\bibitem{MTW} C.~Misner, K.~Thorne, J.~Wheeler, {\em Gravitation}, (Freeman, San Francisco, 1973) Chap. 38 .

\bibitem{dhurandhar+tinto} S.~Dhurandhar and M.~Tinto, Mon. Not. R. Astr. Soc. {\bf 234}, 663 (1988).

\bibitem{helstrom} C.~W.~Helstrom, {\em Statistical Theory of Signal Detection}, 2nd ed.
(Pergamon press, London, 1968).

\bibitem{matlab} Matlab, URL: http://www.mathworks.com.

\bibitem{lazarus}J.~Baker {\em et al}, { Phys. Rev.} D {\bf 65}, 124012 (2002).
\end{thebibliography}
\end{document}